\documentstyle[aps,epsf]{revtex}

\begin{document}
\draft
\title{Quintessence at Galactic Level?}

\author{T. Matos, and F.\ S.\ Guzm\'an\footnote{E-mail:
siddh@fis.cinvestav.mx}}
\address
  {Departamento de F\'{\i}sica,\\
   Centro de Investigaci\'on y de Estudios
   Avanzados del IPN,\\
   AP 14-740, 07000 M\'exico D.F., MEXICO.
  }

\maketitle

%%%%%%%%%%%%%%%%%%%%%%%%%%%%%%%%%%%%%%%%%%%%%%%%%%%%%%%%%%%%%%%%%%%%%%%%%%%%% 
\begin{abstract}
Recently it has been proposed that the main contributor to the dark energy
of the Universe is a dynamical, slow evolving, spatially inhomogeneous
scalar field, called the quintessence. We investigate the behavior of this
scalar field at galactic level, trying it as the dark matter in the halos of
galaxies. Using an exact solution of the Einstein's equations, we find an
excellent concordance between our results and observations.\\
\end{abstract}

Recent observations
in type Ia supernovae have suggesteded the value $\Omega _{0}\sim 1$
considering 
$\Omega _{\Lambda }\sim 0.6$ \cite{perlmutter}. 
From the theoretic point of view
some cosmologists prefer to explain the observed data coming from Ia
supernovae using a decaying cosmological constant \cite{turnerQ}, which
simplest realization is a dynamical, slow evolving, spatially inhomogeneous
scalar field, called the quintessence \cite{stein1}. Quintessence is the
ingredient the Universe should contain, in order to explain the Cosmic
Background Radiation, large scale structure and the cosmic acceleration 
of the Universe \cite{stein1,stein2}. The quintessence is postulated as a
slow varying scalar
field in order to obtain equations of state $p=\omega (t)\rho $, with
$\omega (t)<-0.6$ for this substance. 
Here we investigate further consequences of the scalar
field hypothesis. A fundamental question arises: how does such scalar
field behaves locally? Given that the scalar field varies
slowly in time, one can study the space variation alone in a certain region
of the space-time, neglecting the time variation of the scalar field. Let us
start from the scalar field action

\begin{equation}
S=\int d^{4}x\sqrt{-g}[-\frac{\cal R}{\kappa _{0}}+2(\nabla \Phi
)^{2}-V(\Phi )]
\label{1}
\end{equation}

\noindent 
where ${\cal R}$ is the scalar curvature, $\Phi $ the scalar field,
$\kappa _{0}=8\pi G$ and $V(\Phi )=\Lambda e^{-2\kappa _{0}\Phi }.$ As
an example we present a model of a spiral galaxy by
considering the scalar field of action (\ref{1}) as the dark matter
component, for which we use an exponential potential, as found convenient
for quintessence \cite{turnerQ}. In order to do so we suppose the analog
of quintessence but with the scalar field dependence on the spatial
coordinates. 
\newline

We assume that the space-time of a galaxy has to be axial symmetric and
static; this last condition is reasonable if one considers that dragging
effects are unappreciable given the small pressure between stars. The most
general line element having these properties written in the Papapetrou form
is

\begin{equation}
ds^{2}=\frac{1}{f}[e^{2k}(d\rho^2 + d\zeta^2)+W^{2}d\phi ^{2}]-f\
c^{2}dt^{2},
\label{metric1}
\end{equation}
with the functions $%
f,\ W$ and $k$ depending only on $\rho $ and $\zeta $.
The general field equations obtained from (\ref{1}) are the Klein-Gordon and
the Einstein's equations

\begin{equation}
\Phi _{;\mu }^{;\mu }-\frac{1}{4}\frac{dV}{d\Phi }=0 \ , \  \  \  \  \  \
{\cal R}_{\mu \nu } =\kappa _{0}[2\Phi _{,\mu }\Phi _{,\nu
}+\frac{1}{2}g_{\mu
\nu }V(\Phi )]  \label{Eins}
\end{equation}

\noindent where a semicolon means covariant derivation. 
It has been shown that an exact solution for the
system (\ref{Eins}) is \cite{us},

\begin{equation}
ds^{2}=\frac{r^{2}+b^{2}\cos ^{2}\theta }{f_{0}}(\frac{dr^{2}}{r^{2}+b^{2}}%
+d\theta ^{2})+\frac{r^{2}+b^{2}\sin ^{2}\theta }{f_{0}}d\phi
^{2}-f_{0}c^{2}(r^{2}+b^{2}\sin ^{2}\theta )dt^{2}  \label{metbl}
\end{equation}

\noindent and the effective energy density is

\begin{equation}
\mu _{DM}=V(\Phi )=\frac{4f_{0}}{\kappa _{0}(r^{2}+b^{2}\sin ^{2}\theta )}
\label{mu}
\end{equation}

\noindent
being $\rho =\sqrt{(r^{2}+b^{2})}\sin \theta $, $\zeta =r\cos \theta $
the Schwarzschild-like coordinates.
Solution (\ref{metbl}) is the space-time of a ``scalar field soup'',
therefore it is not asymthotically flat, this means that the dark matter
here behaves in a completely relativistic manner, there is no Newtonian
limit. It is remarkable the coincidence with the profile of an isothermal
halo in the equatorial plane \cite{begeman}. 
The geodesic equations for test particles (stars) into the equatorial
plane of our space-time read

\begin{equation}
\label{chr1}
\frac{\partial ^2 R}{\partial \tau ^2} - R \left(\frac{\partial
\phi}{\partial \tau} \right)^2 + f_{0}^{2} c^2 R \left(\frac{\partial
t}{\partial \tau}\right)^2 = 0, \  \  \
\frac{\partial \phi}{\partial \tau} = \frac{B}{R^2 f_0}, \  \  \
\frac{\partial t}{\partial \tau} = \frac{A}{R^2 f_0}
\end{equation}

\noindent
where $\tau$ is the proper time of the test particle and
$R=\int{ds}=  \sqrt{(r^2+b^2)/f_0}$ is the proper distance of the test
particle at the equator from the galactic center. Observe that for a
circular trajectory, the first of equations (\ref{chr1}) reduces to

\begin{equation}
\label{chr2}
\dot{\phi} = f_0 c = \frac{B}{A}
\end{equation}

\noindent
where the dot means derivative with respect to $t$ and we have used the
other two geodesic equations for the second identity. $A$ and $B$ are two
constants of motion of the test particle we are considering. We can
estimate the constant $A$ using the invariance of the metric. At the
equator it is obtained

\begin{equation}
\label{chr3}
ds^2 = 
-\left(f_0(r^2+b^2) - \frac{v^2}{c^2} \right)c^2
dt^2  = -c^2d\tau^2
\end{equation}

\noindent
with $v^{2}=g_{ij}v^{i}v^{j}$, $v^{i}=(\dot{r},\dot{\theta},\dot{%
\phi})$, 
from where it arises an expression for $A$ in terms of the metric
functions

\begin{equation}
\label{chr4}
A = \frac{r^2+b^2}{\sqrt{f_0(r^2+b^2)-v^2/c^2}} \sim
\sqrt{\frac{r^2+b^2}{f_0}} = R
\end{equation}

\noindent
since $v^2 \ll c^2$. Using (\ref{chr2}) and (\ref{chr4}) we obtain an
estimation for the angular momentum $B =
v_{DM} R$, which implies $B \sim f_0 cR$, and therefore $v_{DM}
\sim f_0 c
=constant$ in the regions where the scalar matter dominates. This
remarkable result qualitatively agrees with observations, it means that
the circular velocity of a star far away from the center of the galaxy
$v_{DM}$ does not depend on the distance $R$. 
Furthermore, the angular momentum of test
particles is determined by the luminous matter near the center of the
galaxy, where $B=f_0 cR \sim 0$. 
In the following we use the approximation $B \sim f_0 cR$
along the whole galaxy.
The first of equations (\ref{chr1}) is the second Newton's law for
particles travelling into the scalar field background. We can interpret

\begin{equation}
\label{chr5}
\frac{\partial^2 R}{\partial \tau^2} = R \left(\frac{\partial
\phi}{\partial \tau} \right)^2 - f_{0}^{2}c^2R\left(\frac{\partial
t}{\partial \tau} \right)^2 = \frac{B^2}{R^3 f_{0}^{2}} - c^2
\frac{A^2}{R^3} = \frac{c^2}{R} - c^2 \frac{A^2}{R^3} 
\end{equation}

\noindent
as the force due to the scalar field background, i.e. $F_{\Phi} =
c^2/R-c^2 A^2/R^3$. We know that the luminous matter is completely
Newtonian. At the other hand, the Newtonian force due to the luminous
matter is given by $F_{L} = GM(R)/R^2 = v_{L}^{2}/R = B_{L}^{2}/R^3$,
where $v_L$ is the circular velocity of the test particle due to the
contribution of the luminous matter 
and $B_L$ is its corresponding angular
momentum per unit of mass. The total force acting on the test particle is
then $F = F_{\Phi} + F_{L}$. For circular trajectories $\partial^2
R/\partial \tau^2 = F = 0$, then

\begin{equation}
\label{chr6}
\frac{B_{L}^{2}}{R^2} - c^2 \frac{A^2}{R^2} = -c^2
\end{equation}

The constants of motion are the total energy per unit of mass of the test
particle and its angular momentum per unit of mass

\begin{equation}
(E/m)^2 = \frac{f_0^2 c^4(r^2 + b^2)^2}{f_0(r^2+b^2) - v^2/c^2}
, \  \  \  \
(l/m)^2 = \frac{v^2(r^2+b^2)}{f_0(f_0(r^2 + b^2) -v^2/c^2)}
\end{equation}

\noindent 
given $l/m=B_L$, and $E/l=c^2f_0 A$. Since $v^{2}\ll c^{2}$
the geodesic equation implies the main result of this work \cite{us}

\begin{equation}
v_{DM}=f_{0}l/m  \label{mean}
\end{equation}

\noindent
i.e. the velocity is independent of the radius as observed at large radii,
where the dark matter dominates. 
This velocity should be the contribution of our scalar dark matter to the
velocity of test particles, and this is why we label it $v_{DM}$.
In order to model completely the rotation curves we introduce a typical
exponential and thin distribution of luminous matter for the disc, for
wich it is useful to consider the Universal Rotation Curve expression
\cite{persic}

\begin{equation}
v_{L}^{2} = v^2(R_{opt})\beta \frac{1.97 x^{1.22}}{(x^2 + 0.78^2)^{1.43}}
\label{vL}
\end{equation}

\noindent
an approximation that works for a sample of 967 spiral galaxies
\cite{persic}. In (\ref{vL}) $x=r/R_{opt}$, the parameter
$\beta=v_{L}(R_{opt})/v(R_{opt})$ being $R_{opt}$ the radius into which it
is contained the 83\% of the onservable mass of the galaxy and $v$ is the
observed circular velocity. Thus we find the
expression $l/m=v_{L}\bar{r}=v_{L}\int {ds}=v_{L}\sqrt{(r^{2}+b^{2})/f_0}$
for the
angular momentum per unit of mass into our space-time. Using this result and
the assumption of a galaxy being a virilized system, the total circular
velocity of a test particle is

\begin{equation}
v_{C}^{2}=v_{L}^{2}+v_{DM}^{2}=v_{L}^{2}(f_{0}(r^{2}+b^{2})+1).  \label{vc}
\end{equation}

%%%%%%%%%%%%%%%%%%%%% FIGURE %%%%%%%%%%%%%%%%%%%%%%%%%%

\begin{figure}[h]
\label{fig1}
\vspace*{-2cm}
\centerline{ 
\epsfxsize=9cm \epsfbox{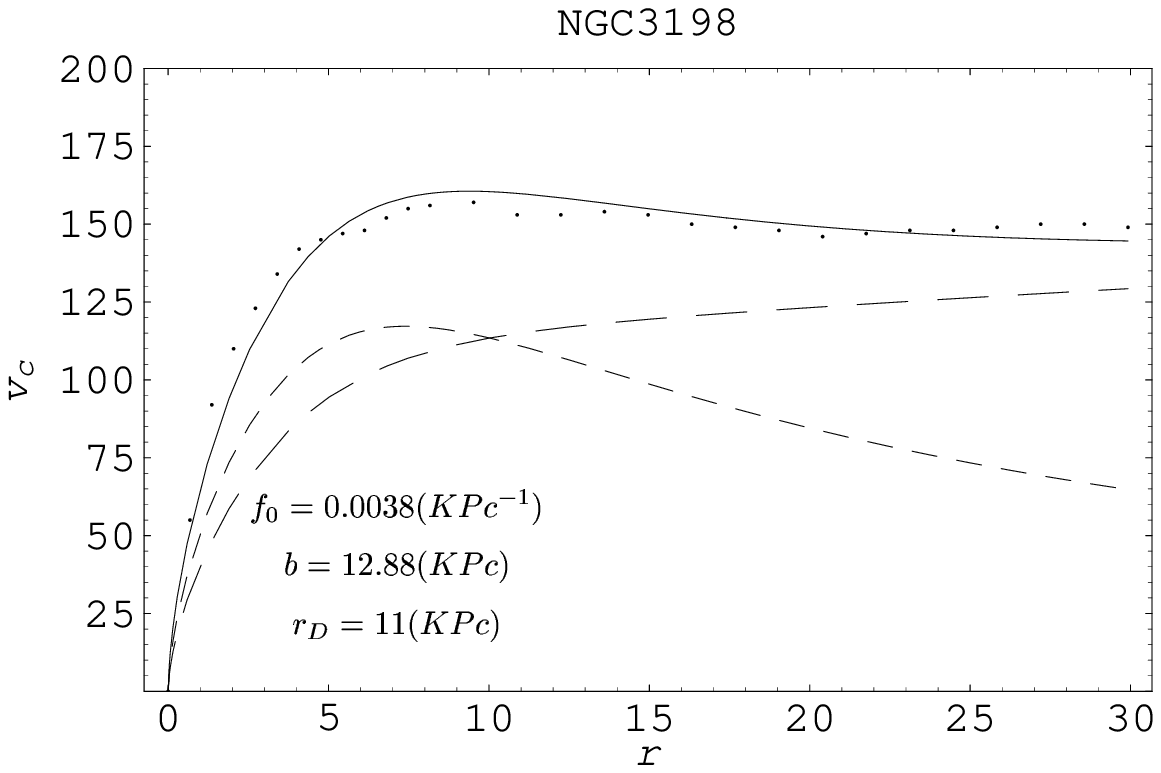} 
\hspace*{-3cm}
\epsfxsize=9cm \epsfbox{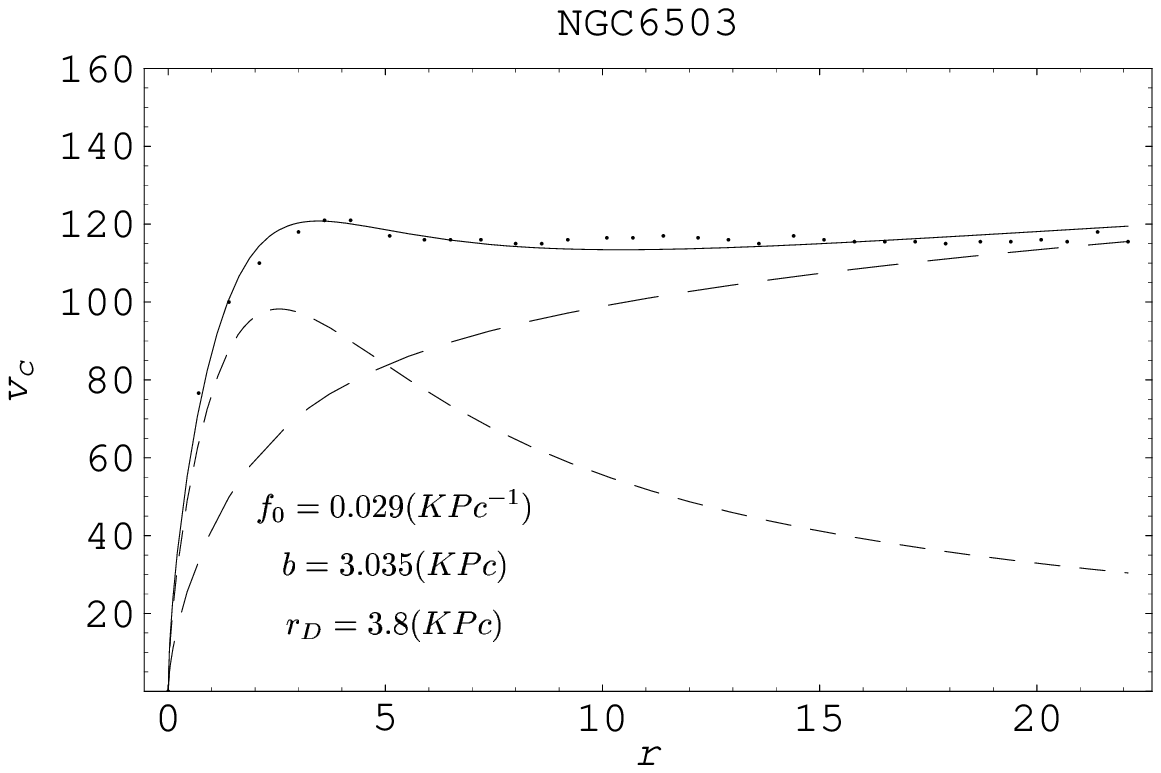}}  
\vspace*{-6.5cm}
\caption
{The circular velocity profiles of four galaxies. 
Continious lines represent the total circular velocity ($v_C$),
long-dashed the contribution of the dark matter to the total velocity
($v_{DM}$) and the short-dashed curves the contribution of luminous
matter ($v_L$); finally the dots represent the observational data. The
value of $r_D$ is the observationally correponding to each of both
galaxies $[$7$]$. 
The units are in (Km/s) in the vertical axis and
in (Kpc) in the horizontal one.}
\end{figure}

The comparison of this model with experimental results of two spiral
galaxies is shown in Fig. 1, where it is evident the great coincidence of
our results with the observed rotation curves. 
Observe that except the relation $v^2 \ll c^2$, our main result
(\ref{mean}) is exact, and that the contribution of luminous matter to the
circular velocity (\ref{vL}) is a very convincing phenomenological model.
This result puts the scalar field as a good candidate to be the dark
matter in the halos of galaxies. It is remarkable that if this result is
confirmed in some way, the scalar field could represent 35\% of dark
matter and 60\% of dark energy in the Universe, it means that the
scalar field could be 95\% of the matter in the whole Universe as
suggested in \cite{luis}.

\acknowledgements{
We want to thank the relativity group in Jena fot its kind hospitality.
This work was partly supported by CONACyT M\'{e}xico, grant 3697-E.}

\end{document}